\input Tex-document.sty

\def\volumeno{I}
\def\firstpage{495}
\def\lastpage{504}
\pageno=495

\def\shorttitle{Singularities in String Theory}
\def\shortauthor{E. Witten}

\title{\centerline{Singularities in String Theory}}

\author{E. Witten\footnote{\eightrm *}{\eightrm Institute For Advanced Study,
Princeton NJ 08540, USA. E-mail: witten@sns.ias.edu}}

\vskip 7mm

\centerline{\boldnormal Abstract}

\vskip 4.5mm

{\narrower \ninepoint \smallskip String theory is a quantum theory
that reproduces the results of General Relativity at long
distances but is completely different at short distances.
Mathematically, string theory is based on a very new --- and
little understood --- framework for geometry that reduces to
ordinary differential geometry when the curvature is
asymptotically small. In the 1990's, many interesting results were
obtained about the behavior of string theory in spacetimes that
develop singularities.  In many cases, the physics at the
singularity is governed by an effective Lagrangian constructed
using an interesting bit of classical geometry such as the
association of A-D-E groups with certain hypersurface
singularities or the ADHM construction of instantons. In other
examples, the physics at the singularity cannot be described in
classical terms but involves a non-Gaussian conformal field
theory.

\def\,{{\hskip .2 cm}}

\vskip 4.5mm



}

\vskip 10mm

\head{1. Introduction}

The classical Einstein equations $$R_{IJ}=0,$$ where $R$ is the
Ricci tensor of a metric $g$ on spacetime, are scale-invariant. In
other words, they are invariant under the scaling of the metric
$g\to tg$, with $t$ a real number; the Ricci tensor is invariant
under this scaling.  A quantum theory of gravity, however, cannot
have this symmetry, since quantum theory depends on the {\it
action}, not just the field equations, and the Einstein-Hilbert
action $$I={1\over 16 \pi G}\int d^nx \sqrt g R$$ ($G$ is Newton's
constant and $n$ is the dimension of spacetime) is not invariant
under scaling.    In fact, under $g\to tg$, the action scales as
$I\to t^{n/2-1}I$, and so is not scale-invariant in any dimension
above two.  (Two dimensions --- a case that figures in string
theory --- is completely special as $I$ is a topological
invariant, the Gauss-Bonnet integral.)

Generally speaking, the classical limit of quantum mechanics
arises by making a stationary phase approximation to a function
space integral.  That integral is very roughly of the form $$\int
D\Phi \,\,\exp(iI/\hbar),$$ where the integral runs over all
fields $\Phi$.  For example, for General Relativity, $\Phi$ would
be a metric tensor on spacetime, perhaps together with other
fields, and $I$ would be the Einstein-Hilbert action defined in
the last paragraph, together with a suitable action for the other
fields, if any.  Here $\hbar$ is Planck's constant, and the
stationary phase approximation to the function space integral is
valid when the appropriate $I/\hbar $ is large.  Note that in
classical physics, $I$ is not a dimensionless number but has
units of ``action'' or energy times time; it is only with the
passage to quantum mechanics that there is a natural constant of
action, namely $\hbar$, and it makes sense to say that in a given
physical situation the action is large or small.

As an example of this criterion, consider black holes.  A
classical black hole can, because of the scale invariance, have
any possible mass $M$ or radius $r$; in four dimensions, for
example, the mass and radius are related (for neutral, unrotating
black holes) by $M=rc^2/G$, where $c$ is the speed of light.  The
value of $I/\hbar$ integrated over the relevant time, which is the
time for light to cross the black hole, is $${I\over
\hbar}=Mc^2\cdot {r\over c}\cdot {1\over \hbar}={GM^2\over \hbar
c}={r^2c^3\over G\hbar}.$$ The classical description of black
holes is valid if this expression is large, or in other words if
$M>>10^{-5}$ gm  or $r>>10^{-33}$ cm.

Known astrophysical black holes have masses comparable to that of
the sun (about $10^{33}$ gm) and above, so unless we get lucky
with mini-black holes left over from the Big Bang, we are not
going to be able to observe what happens for black holes so light
and small that the classical description fails.  But curiosity
compels us to ask how to describe a small black hole, or what
happens near the Big Bang, where classical General Relativity
breaks down for similar reasons to what I have just described for
Black Holes.

Here we run into a problem.  One can read a textbook recipe for
quantization in Dirac's old book or in more modern texts on
quantum field theory.  But these recipes, applied to the
Einstein-Hilbert theory, do not work.   Because of the highly
nonlinear nature of Riemannian geometry, these methods fail to
give a consistent and meaningful result.

This problem is very hard to convey to a mathematical audience
because the whole question is about quantum field theory, which
is not in clear focus as a mathematical subject.  The foundation
for our modern understanding of elementary particles and forces
is the success in quantizing theories such as Yang-Mills theory,
whose action is $$I={1\over 4e^2}\int {\rm Tr}\,F\wedge *F,$$
where $F$ is the curvature of a connection,  $e$ is a real
constant, known as the gauge coupling constant, and ${\rm Tr}$ is
an invariant quadratic form on the Lie algebra of the gauge group.
Other somewhat analogous theories such as the quantum theory of
maps from a Riemann surface to a fixed Riemannian manifold have
also been extensively explored by physicists, with applications
to both string theory and condensed matter physics.   Apart from
their central role in physics, quantum gauge theory and its
cousins are the basis for the application of physical ideas to a
whole range of mathematical problems, from the Jones polynomial
to Donaldson theory and mirror symmetry.

But the understanding of quantum field theory that physicists
have gained, though  convincing and sufficient to make many
computations possible, is hard to formulate rigorously.  This
makes it difficult for mathematicians to understand the questions
of physicists, much less the partial results and approaches to a
solution. In ``constructive field theory,'' some of the standard
physical claims about quantum field theory have been put on a
rigorous basis, but there is still a long way to go to
effectively bridge the gap.

For physicists, quantum gravity is an important problem not only
because we would like to understand black holes, the Big Bang, and
the quantum nature of spacetime --- but also because reconciling
General Relativity and quantum mechanics is necessary if we are to
unify the forces of nature.  We cannot achieve a unified
understanding of nature if gravity is understood one way and the
subatomic forces are understood by a different and incompatible
theory.  Moreover, the difficulty in reconciling these theories is
probably our best clue about understanding physics at a much
deeper level than we understand now.  So what has been
accomplished toward reconciling General Relativity and quantum
mechanics?

\head{2. String theory}

Not much has been learned by direct assault.  But roughly thirty
years ago, in trying to solve another problem, physicists stumbled
in ``string theory'' onto a very rich and surprising new
framework for physics and geometry, which apparently does yield a
theory of quantum gravity, though we do not understand it very
well yet.  String theory introduces in physics a new constant
$\alpha'\sim (10^{-32}\,{\rm cm})^2$ (read ``alpha-prime''), which
is somewhat analogous to Planck's constant $\hbar\sim
10^{-27}\,{\rm erg\,sec}$, and modifies the concepts of physics in
an equally far-reaching way.

If string theory is correct, then both $\hbar$ and $\alpha'$ are
nonzero in nature.  The deformation of classical physics in
turning on nonzero $\hbar$ is comparatively familiar to most in
this audience, at least at the level of nonrelativistic quantum
mechanics (as opposed to quantum field theory): classical
concepts such as the position and velocity of a particle become
``fuzzy'' in the transition to quantum mechanics.  Turning on
$\alpha'\not= 0$ introduces an additional fuzziness in physics,
roughly as a result of turning particles into strings.  One aims
to unify the forces by interpreting all of the different
particles in nature as different vibrational states of one basic
string.

String theorists spend the late 1980's and early 1990's largely
studying the $\alpha'$ deformation.  This work is hard to
describe mathematically because it is all based on techniques of
quantum field theory.  Roughly it involves a new kind of geometry
in which one is not allowed to talk about points or geodesics but
one can talk about (quantum) minimal surfaces.  This blurs all
the classical concepts in geometry and makes possible nonclassical
behavior.  The new fuzziness has a characteristic scale
$\sqrt{\alpha'}\sim 10^{-32}\,{\rm cm}$, and has many
consequences. On much larger scales, just like the quantum
uncertainty in gravity, the stringy fuzziness is unimportant. But
it is important if one looks closely and can lead to nonclassical
behavior, such as mirror symmetry.

A commonly encountered framework for nonclassical behavior in
string theory is the following. Consider a family of classical
solutions of string theory depending on several parameters; call
the parameter space ${\cal N}$.  At a generic point in ${\cal N}$,
the stringy effects are important and one cannot usefully
describe the situation in terms of a classical spacetime.  As one
approaches a special point $P\in {\cal N}$ (or more generally
some locus in ${\cal N}$ of positive codimension), the relevant
length scales in spacetime become large, the string effects
become unimportant, and a classical spacetime $X$ emerges.  ($P$
is typically a cusp-like point in ${\cal N}$; that is, there is
typically a natural metric on ${\cal N}$, and $P$  is at infinite
distance in this metric.)  In that limit, the string equations
reduce to the classical Einstein equations on $X$ (or more
precisely, their appropriate supersymmetric extension, about
which more later). To run what I have said so far in reverse,
starting with a classical spacetime $X$ that is embedded in
string theory, if the radius of $X$ (and every relevant length
scale) is large compared to $\sqrt{\alpha'}$, then classical
geometry is a good approximation to the stringy situation.  But
by varying the parameters so that the ``radius''
 of $X$ is not large compared to $\sqrt{\alpha'}$, one can get a
situation in which classical geometry is not a good approximation
and must be replaced with stringy geometry. This situation is
described by points in the interior of ${\cal N}$.

A more seriously nonclassical behavior arises if in addition to
the point $P$,  there are additional points $Q,R\in {\cal N}$ at
which classical behavior again arises, but this time with
different classical spacetimes $Y,Z$ of different topology.  In
this case, by moving in ${\cal N}$ from $P$ to $Q$, we can go
smoothly from a situation in which classical geometry is a good
approximation and the spacetime is $X$, to a situation in which
classical geometry is again a good approximation but the
spacetime is $Y$.   For this process to occur smoothly even
though the initial and final spacetimes have different topology,
it inevitably happens that in interpolating from $P$ to $Q$, one
has to pass through a region in which classical geometry is not a
good approximation.

The example of what I have just described that is most discussed
mathematically is that in which $X$ is a Calabi-Yau threefold,
and, say,  $Y$ is mirror to $X$ and $Z$ is another Calabi-Yau
manifold that is  birational to $X$ or $Y$.

Because the characteristic length scale of stringy behavior, in
the simplest way of matching string theory with the real world,
is about $10^{-32}$ cm, way below the distance scale that we can
probe experimentally, much of the structure of string theory,
assuming it is right, is out of reach experimentally.
Conceivably, we might one day be able to use string theory to
calculate masses and interaction rates of the observed elementary
particles, but this seems far off.  It is also just possible that
we might have the chance to observe one or another kind of
massive string relic left over from the early universe. There is,
however, another possibility that is much more likely for the
immediate future; there is one important aspect of stringy
geometry which may very well be accessible to experiments.  This
is ``supersymmetry,'' roughly the notion that spacetime is more
accurately understood as a supermanifold, with both odd and even
coordinates, rather than as an ordinary manifold.  The theory of
supermanifolds is more accessible and better known mathematically
than some of the other things that I have mentioned.  Most of the
known geometrical applications of quantum field theory involve
supersymmetry in one way or another.

If supersymmetry is relevant in nature, the oscillations of known
particles in the ``odd'' directions in spacetime would give new
elementary particles that could be discovered in accelerators.
There are hints that these new particles exist at energies very
close to what has already been reached experimentally.   To me the
most striking hint of this comes from the measured values of the
strong, weak, and electromagnetic coupling constants, which are
in excellent agreement with a prediction based on supersymmetric
grand unification. If these hints have been correctly
interpreted, we are likely to discover supersymmetric particles at
accelerators in this decade, probably at the Fermilab accelerator
in Illinois or at the Large Hadron Collider, which is being built
at the European laboratory CERN near Geneva.

It is interesting to contemplate the impact on mathematics if
supersymmetry is really discovered experimentally.  When General
Relativity emerged as an improvement on Newton's theory of
gravity, this gave a huge boost to the mathematical investigation
of Riemannian geometry.  Nonrelativistic quantum mechanics
probably gave an equivalent boost to functional analysis. Quantum
field theory is so multi-faceted that a simple summary of its
mathematical influence is difficult; some aspects of quantum field
theory have influenced mathematics considerably, but as I have
explained, problems of rigor have kept the core ideas of this
richest of physical theories inaccessible mathematically.
Supersymmetry, I think, would fall somewhere in between.  Its
experimental discovery would greatly increase the interest of
mathematicians in supermanifold theory, which is accessible
mathematically, but the full impact on mathematics would be
delayed because the real payoff of supersymmetry lies in the
realm of quantum field theory and string theory.

\eject

\head{3. Uniform breakdown of the ``large, smooth'' approximation}

In recent years, the most significant development in string theory
has been to understand some of the things that happen when both
$\hbar$ and $\alpha'$ are important.  One discovery is that the
different models of string theory that we knew of in the 1980's
are related like the different classical spacetimes $X,Y$, and
$Z$ that we discussed above.  In an asymptotic expansion near
$\hbar=0$, they are different, but in a more complete
description, the different string models arise as different
semiclassical limits of one richer theory that has been dubbed
$M$-theory.  $M$-theory, though we do not really understand it
yet, is thus the candidate for super-unification of the laws of
nature.

Also, we have gained insight about what happens in many
situations in which classical geometry breaks down.  In the
``large, smooth'' limit of spacetime, in which all relevant
length scales are large, classical geometry (enriched to include
supersymmetry) is always a good approximation.  But what happens
when the ``large, smooth'' approximation breaks down?

\def\S{{\bf S}}
\def\K3{{\rm K3}}
\def\I{{\bf I}}
\def\T{{\bf T}}

Many interesting results were obtained in the 1990's about
situations in which the ``large, smooth'' approximation breaks
down everywhere at once.  I will give a few  examples. These
examples involve the basic ten-dimensional models of string
theory, such as Type IIA and Type IIB superstrings and the
heterotic string, and also the eleven-dimensional $M$-theory. Let
$\S^1(r)$ denote a circle of circumference $2\pi r$. Then our
first example is the assertion for any ten-dimensional spin
manifold $X$, Type IIA superstring theory on $X\times \S^1(r)$ is
equivalent to Type IIB superstring theory on $X\times
\S^1(\alpha'/r)$. If $r>>(\alpha')^{1/2}$, the description via
Type IIA superstring theory is  transparent as ordinary
geometrical concepts are valid, while for small $r$ the second
description is better. Starting on the Type IIA side at large
$r$, the ``large, smooth'' description breaks down for $r\to 0$
(as there are closed geodesics of length $2\pi r$ in $X\times
\S^1(r)$), and the equivalence to Type IIB on $X\times
\S^1(\alpha'/r)$ gives a description that is valid when the Type
IIA description has failed.

My other examples will relate an eleven-dimensional description
via $M$-theory to a ten-dimensional string theory. With $X$ as
before a ten-dimensional spin manifold and $Y$ a seven-dimensional
spin manifold, and letting $\K3(r)$ denote a K3 surface of radius
$r$ and $\I(r)$ a length segment of length $r$, and $\T^3$ a
three-torus, we have the following relations: (i) $M$-theory on
$X\times \S^1(r)$ is equivalent as $r\to 0$ to Type IIA
superstring theory on $X$; (ii) $M$-theory on $Y\times \K3(r)$ is
equivalent as $r\to 0$ to the heterotic string on $Y\times \T^3$;
and $M$-theory on $X\times \I(r)$ is equivalent for $r\to 0$ to
the $E_8\times E_8$ heterotic string on $X$. In each of these
examples, the ``large, smooth'' approximation is valid for large
$r$ (if $X$ and $Y$ are large enough) and breaks down for small
$r$. In each  example, the string coupling constant in the string
theory description vanishes for $r\to 0$, so that the string
theory description is useful in that limit --- an asymptotic
expansion valid for small $r$ can be explicitly worked out, giving
a detailed answer to the question of what happens when the
``large, smooth'' approximation fails. Note that these examples
involve highly nonclassical behavior, with change in the topology
and even the dimension of spacetime --- for example,  a
four-dimensional K3 surface at large $r$ is replaced by a
three-dimensional torus when $r$ becomes small.

Relations of this type are ``quantum'' analogs (involving both
$\hbar$ and $\alpha'$) of mirror symmetry (which from this
standpoint involves only $\alpha'$) and have led to the
understanding that the different string models are different
limits of the same things.  Many other examples have been worked
out in which the ``large, smooth'' approximation breaks down
everywhere at once.  I want to focus in the remaining time today,
however, on another type of situation.  This is the case in
which, as some parameter is varied, the ``large, smooth''
approximation remains valid generically, but breaks down along
some locus of codimension $d>0$ where spacetime develops a
conical singularity.

\head{4. Behavior at conical singularities}

The behavior of string theory when spacetime develops a conical
singularity in positive codimension can be investigated by
methods that exploit the fact that the ``large, smooth''
approximation remains generically valid, away from the
singularity.  One often can identify an interesting mathematical
and physical phenomenon supported at the singularity.  I will
select examples in which both $\hbar$ and $\alpha'$ play an
important role. There also are many instances of conical
singularities that can be studied at $\hbar=0$, such as the
string theory orbifolds that have motivated one of the satellite
meetings of ICM-2002. But we will focus on problems that involve
both $\alpha'$ and $\hbar$.  I will give three examples; two
involve known mathematical constructions that appear in a new
situation, while in the third the key phenomenon is nonclassical
--- it can only be formulated quantum mechanically.

\def\R{{\bf R}}
\def\C{{\bf C}}
 {\bf I.} {\it $M$-Theory At An A-D-E Singularity:}   The A-D-E singularities
are codimension four singularities that look locally like
$\R^4/\Gamma$, where $\Gamma$ is a finite subgroup of $SU(2)$,
acting on $\R^4\cong \C^2$ and preserving the hyper-Kahler
structure of $\R^4$.  An extensive mathematical theory relates the
A-D-E singularity to the A-D-E Dynkin diagram and many associated
bits of geometry and algebra.  However, the role of the A-D-E {\it
group} in relation to the singularity is elusive.   Like other
singular spaces, the A-D-E singularity is usefully studied as a
limit of smooth spaces carrying the appropriate structure. In this
example, the singular space $\R^4/\Gamma$ has a hyper-Kahler
resolution (due to Kronheimer) that contains exceptional divisors
of area $A_1,\dots, A_r$ which appear as parameters in the metric
($r$ is the rank of the relevant A-D-E group, and the intersection
form of the divisors is minus the Cartan matrix of the group).  In
the context of $M$-theory or string theory, for $A_1,\dots,A_r$
large, the ``large, smooth'' approximation is valid and classical
geometry can be applied.  We want to know what happens as
$A_1,\dots,A_r\to 0$, giving a singularity at the origin in
$\R^4$.  This is the basic A-D-E singularity in $\R^4$; in
eleven-dimensional $M$-theory, we would usually be working on an
eleven-dimensional spacetime $X$, and the singularity arises on a
codimension-four submanifold $Q$. The answer has turned out to be
that in this limit, A-D-E gauge fields --- and their
supersymmetric extension --- appear on $Q$. Assuming that the
``large, smooth'' approximation has failed only because of the
A-D-E singularity, there is an effective description of the
resulting physics that roughly speaking is governed by the action
$$I=\int_Xd^{11}x\sqrt g\left(R+\dots\right) +\int_Q{\rm
Tr}\,\left( F\wedge *F+\dots\right),$$ where $R$ is the Ricci
scalar, $F$ is the curvature of the A-D-E connection, and
``$\dots$'' refers to the supersymmetric extension.  The
supersymmetric extension of the gauge theory that is supported on
$Q$ turns out to automatically contain the variables needed to
parametrize Kronheimer's hyper-Kahler resolution of the
singularity.

{\bf II.} {\it Gauge Theory Instantons:}   My second example
involves the instanton solutions of four-dimensional Yang-Mills
theory.  Like the Einstein equations, the equations for
Yang-Mills instantons, which read $F=-*F$, where $F$ is the
curvature of a Yang-Mills connection on $\R^4$, are scale
invariant.  (In fact, they have the much stronger property of
conformal invariance.)  Instantons therefore come in all sizes.
One can scale the size of an instanton all the way down to zero,
giving, in the limit, a singular, point-like instanton.  So (even
on a compact four-manifold) instanton moduli space is
non-compact. This raises the question, ``What happens when an
instanton becomes small?''

The answer to this question depends on what one is trying to do. I
will describe three possible answers.  {\it (i)}  Instantons were
introduced in quantum field theory in the mid-1970's.
Traditionally, physicists were interested in certain integrals
over instanton moduli space.  From this point of view, the meaning
of the noncompactness of instanton moduli space is clear: one
should make sure that the integrals of interest converge, and one
should be careful when integrating by parts. {\it (ii)} In
Donaldson theory, one is interested in intersection theory on
instanton moduli space; ``instanton bubbling'' --- the shrinking
of an instanton to a point --- is the main source of technical
difficulty.  One deals with it by a variety of technical means
such as considering  cycles in moduli space whose intersections
avoid the bubbling region.  {\it (iii)} In string theory, one
expects the classical instanton equation $F=-*F$ to be a good
approximation for large instantons, this being an example of the
validity of classical concepts in the ``large, smooth'' region.
But one expects this description  to break down as the instanton
shrinks. The question here is to find a description that is valid
for small instantons.

The instanton problem can be embedded in string theory in
different ways, so there are several answers.   I will give the
answer in one case --- Type I superstring theory or the $SO(32)$
heterotic string.  (At the very end of this talk, I briefly point
out a second case.)

Before going on, I should mention one surprising part of the
mathematical theory of instantons.  This is the ADHM construction
(due to Atiyah, Drinfeld, Hitchin, and Manin) of instantons in
$\R^4$.  To describe a $k$-instanton solution of $SU(N)$ gauge
theory on $\R^4$, the ADHM construction employs an auxiliary
$U(k)$ group. (For example, the instanton moduli space is
constructed as a hyper-Kahler quotient of a linear space divided
by $U(k)$.)  The interpretation of this group is somewhat
mysterious in classical geometry, just at the role of the A-D-E
group in relation to the A-D-E singularity is somewhat mysterious
classically.

Instanton bubbling occurs at a point in $\R^4$, so in  gauge
theory in any dimension, it occurs on a submanifold of
codimension four. In ten-dimensional Type I superstring theory on
a ten-manifold $X$, the small instanton thus appears on a
codimension four submanifold $Q$. The answer to the small
instanton problem turns out to be that the $U(k)$ group of the
ADHM construction appears as a gauge group in the physics on $Q$.
The effective action that governs this situation turns out to be
schematically $$I=\int_Xd^{11}x\sqrt g\left(R+\dots\right)
+\int_Q{\rm Tr}\,\left( F\wedge *F+\dots\right),$$ where in this
case $F$ is the curvature of a $U(k)$ connection, while
``$\dots$'' refers to additional terms required by supersymmetry
plus the additional variables used in the ADHM construction to
describe the moduli space as a hyper-Kahler quotient.

So once again, an interesting and surprising bit of classical
mathematics becomes important near the singularity.  I move on
now, however, to an example in which the key phenomenon cannot be
described in classical terms.

{\bf III.} {\it Type IIB At An A-D-E Singularity:}  Here we
consider again the A-D-E singularity, but now in Type IIB
superstring theory rather than in $M$-theory.  The answer turns
out to be completely different: we do not get a description with
new classical degrees of freedom; instead a ``non-trivial'' or
``non-Gaussian'' conformal field theory is supported on the locus
$Q$ of the A-D-E singularity.  The assertion that this theory is
``non-trivial'' means that it exists as a conformally invariant
quantum field theory, but cannot be conveniently described in
terms of classical or Gaussian fields.

This particular nontrivial conformal field theory might be
described as a ``nonabelian gerbe theory''; it is related to
two-forms in roughly the way that nonabelian gauge theory is
related to one-forms.  Classically, one-forms have a nonabelian
generalization in gauge theory, but to find an analogous theory
for two-forms one must apparently go to quantum theory. The
existence and basic properties of this particular six-dimensional
conformal field theory can be used to deduce Montonen-Olive
duality of quantum Yang-Mills theory in four dimensions, and this
in turn has implications for certain four-manifold invariants.

So this example again involves interesting mathematics, but to
describe the result requires use of quantum concepts in a more
intimate way.  The same happens if we consider the small instanton
problem in the $E_8\times E_8$ heterotic string (rather than Type
I or the $SO(32)$ heterotic string as considered above).  There
is no ADHM construction for $E_8$ instantons, so there is no
candidate for an answer along the lines sketched above for Type I;
instead, a non-Gaussian conformal field theory appears on the
small instanton locus $Q$.

A general orientation to the subject matter discussed in this lecture
can be found in the second half of volume 2 of [1]. A few of the
original research papers of relevance are [2] for the A-D-E
singularities, [3] for the ordinary double point singularity in complex
dimension three (we have not actually discussed this case in the present
lecture, but it was important in the development of the ideas), and [4]
for small instantons.  In addition, I have discussed the small instanton
problem from a different but related point of view in [5].  Some readers
may also want to consult general expositions of quantum field theory,
such as the recent textbook [6] for physicists, or the exposition aimed
at mathematicians in [7].  Finally, a comparatively recent account of
known rigorous results on quantum field theory can be found in [8].

\head{References}

\item{[1]} J. Polchinski, {\it String Theory}, Cambridge
University Press.

\item{[2]} E. Witten, String Theory Dynamics In Various
Dimensions, {\it Nuclear Physics}, B443 (1995), 85.

\item{[3]} A. Strominger, Massive Black Holes and Conifolds In
String Theory, {\it Nuclear Physics}, B451 (1995), 96.

\item{[4]} E. Witten, Small Instantons In String Theory,
{\it Nuclear Physics}, {\it Nuclear Physics}, B460 (1996), 541.

\item{[5]} E. Witten, Small Instantons In String Theory, in
{\it Prospects In Mathematics}, ed. H. Rossi, American
Mathematical Society (1999), 111.

\item{[6]} S. Weinberg, {\it The Quantum Theory Of Fields},
Cambridge University Press.

\item{[7]} P. Deligne et. al, eds., {\it Quantum Fields and
Strings: A Course For Mathematicians}, American Mathematical Society
(1999).

\item{[8]} V. Rivasseau, {\it From Perturbative To Constructive
Renormalization}, Princeton University Press (1991).

\end